\documentstyle[prd,aps,preprint,epsfig]{revtex}

\tighten

\begin{document}
\draft
 
\pagestyle{empty}

\preprint{
\noindent
\hfill
\begin{minipage}[t]{3in}
\begin{flushright}
LBNL--42731 \\
UCB--PTH--99/01 \\
January 1999
\end{flushright}
\end{minipage}
}

\title{A large final-state interaction in the $0^-0^-$ decays of $J/\psi$}

\author{
Mahiko Suzuki
}
\address{
Department of Physics and Lawrence Berkeley National Laboratory\\
University of California, Berkeley, California 94720
}

\maketitle

\begin{abstract}
In view of important implications in the $B$ decay, the $0^-0^-$ 
decay modes of $J/\psi$ are analyzed with broken flavor SU(3) symmetry 
in search for long-distance final-state interactions. If we impose one
mild theoretical constraint on the electromagnetic form factors, we find 
that a large phase difference of final-state interactions is strongly 
favored between the one-photon and the gluon decay amplitudes. 
Measurement of $e^+e^-\rightarrow\gamma\rightarrow \pi^+\pi^-$ and
$e^+e^-\rightarrow\gamma\rightarrow K^+K^-$ off the $J/\psi$
peak can settle the issue without recourse to theory.

\end{abstract}
\pagestyle{plain}
\narrowtext

\setcounter{footnote}{0}

\section{Introduction}
     The final state interaction (FSI) in the nonleptonic 
$B$ decay has been an important unsolved issue in connection with
the search of direct CP violations. Unlike the short-distance FSI, 
the long-distance FSI has not been understood well enough
even qualitatively. The experimental data of the $D$ 
decay clearly show that the FSI phases are large in the 
$D\rightarrow \overline{K}\pi$ decay modes\cite{bishai}. 
Opinions divide as to how strong the FSI is in the $B$ decay. 
Some theorists have suggested that the long-distance FSI 
should be small at the mass scale of the $B$ meson. But others have 
obtained large FSI phases by numerical computations based on various 
dynamical assumptions and approximations.  According to the 
latest data, the FSI phases are tightly bounded for $B\rightarrow
\overline{D}\pi$ and a little less so for $B\rightarrow\overline{D}\rho$, 
$\overline{D}^*\pi$ and $\overline{D}^*\rho$\cite{suzuki1}.  However, 
the tight bounds are closely tied to smallness of the so-called 
color-suppressed modes. Is the smallness of the FSI phases 
special only to those sets of modes for which the color suppression 
occurs ? If it is more general, where does transition 
occur from large FSI phases to small FSI phases in terms of the 
mass scale of a decaying particle ?

Although the process is not a weak decay, the $J/\psi$ decay falls 
between the $D$ decay and the $B$ decay in terms of energy scale. Since the 
time scale of the strong and electromagnetic decay processes of $J/\psi$ is
much shorter than that of the long-distance FSI, the decay interactions of
$J/\psi$ act just like the weak interactions of the $D$ and 
the $B$ decay as far as the long-distance FSI is concerned.
For this reason, analysis of the $J/\psi$ decay amplitudes provides 
one extrapolation from the $D$ mass toward the $B$ mass.
    Among the two-body decay modes of $J/\psi$, most extensively measured 
are the $1^-0^-$ modes. A detailed analysis of those decay amplitudes
with broken flavor SU(3) symmetry found a large relative phase of FSI 
($\simeq 75^{\circ}$) between the one-photon and the gluon 
decay amplitudes\cite{suzuki2}. Since there are many independent 
SU(3) amplitudes for the $1^-0^-$ decay, the analysis involved one 
assumption of simplification on assignment of the FSI phases. 

In this short paper, we shall study the $0^-0^-$ decay 
modes of $J/\psi$ which are much simpler in the SU(3) 
structure. The result of analysis turns out clearer 
and more convincing. Once the asymptotic behavior of the 
electromagnetic form factors is incorporated in analysis, 
the current data favor a very large FSI phase difference between 
the one-photon and the gluon decay amplitudes.

\section{Final state interaction}

In order to formulate the FSI, it is customary to separate interactions 
into three parts, the decay interaction, the rescattering interaction,
and the hadron formation interaction. Separation between the second and 
the third can be done only heuristically at best, not at the level of 
Lagrangian. One way to circumvent this ambiguity and see general properties
of the FSI is to break up decay amplitudes in the eigenchannels of the 
strong interaction S-matrix:
\begin{equation}
     \langle\beta|S|\alpha\rangle = 
  \delta_{\alpha\beta}e^{2i\delta_{\alpha}}. \label{eigenchannel}
\end{equation}
An observed two-body final state can be expanded in the eigenchannels 
with an orthogonal matrix as
\begin{equation}
      |ab^{in}\rangle =\sum_{\alpha}O_{ab,\alpha}|\alpha^{in}\rangle , 
                 \label{transf}
\end{equation}
where the superscript ``in'' stands for the incoming state. 
In terms of the ``in'' and ``out'' states, the S-matrix of 
Eq.(\ref{eigenchannel}) can be expressed as $\langle\beta|S|\alpha\rangle
=\langle\beta^{out}|\alpha^{in}\rangle$. When the effective decay 
interactions ${\cal{O}}^{(i)}$, in which we include all 
coefficients, are time-reversal invariant, the decay 
amplitude for $J/\psi\rightarrow ab$ is given in the form 
\begin{equation}
    M(J/\psi\rightarrow ab) = 
     \sum_i\sum_{\alpha}O_{ab,\alpha} M_{\alpha}^{(i)} e^{i\delta_{\alpha}}, 
                  \label{general}
\end{equation}
where $M_{\alpha}^{(i)}e^{i\delta_{\alpha}}$ is the decay amplitude 
into the eigenchannel $\alpha$ through ${\cal{O}}^{(i)}$;
\begin{equation}
   M_{\alpha}^{(i)}e^{i\delta_{\alpha}}=
     \langle ab^{out}|{\cal{O}}^{(i)}|J/\psi\rangle ,
\end{equation}
and $M^{(i)}_{\alpha}$ is real.\footnote{If gluon loop corrections are 
made and analytically continued to the timelike region, ${\cal{O}}^{(i)}$ 
contains a short-distance FSI phase, which is transferred 
into $\delta_i$ in Eq.(\ref{6}).} Two interactions are relevant 
to the $J/\psi$ decay.  For the one-photon annihilation,
${\cal{O}}^{(1)}\propto J_{em}^{\mu}\psi_{\mu}$, where $\psi_{\mu}$ is the 
vector field of $J/\psi$.  For the gluon annihilation,
\begin{equation}
        {\cal{O}}^{(2)} =  F_{\mu}(G)\psi^{\mu},
\end{equation}
where $F_{\mu}(G)$ is a vector function of the gluon field tensor
$G_{\lambda\kappa}$ and its derivatives which is calculated in 
perturbative QCD. When the terms from the same decay interaction are grouped 
together, Eq.(\ref{general}) takes the form,
\begin{equation}
    M(J/\psi\rightarrow ab) = \sum_{i=1,2} M^{(i)}_{ab}e^{i\delta_i},\label{6}
\end{equation}
where
\begin{equation}
         M^{(i)}_{ab}e^{\delta_i}
     = \sum_{\alpha}O_{ab,\alpha}M_{\alpha}^{(i)}e^{i\delta_{\alpha}},     
\end{equation}
We emphasize here that the net FSI phase $\delta_i$ of 
$M^{(i)}_{ab}$ depends on ${\cal{O}}^{(i)}$ through $M_{\alpha}^{(i)}$
even for the same state $ab$ when more than one eigenchannel is open. 
Specifically in the $J/\psi$ decay, $\delta_i$ is different 
between the one-photon and the three-gluon decay amplitude even 
for the same isospin state. If the FSI is strong in the $J/\psi$ decay, 
a large phase difference $\Delta =\delta_1-\delta_2$ can arise. Our aim 
is to learn about $\Delta$ from the decay $J/\psi\rightarrow  0^-0^-$.

\section{Parametrization of amplitudes}   

  One feature of the $J/\Psi\rightarrow 0^-0^-$ is
particularly advantageous to our study: There 
is no SU(3) symmetric decay amplitude for the gluon 
decay. Charge conjugation does not allow a $0^-0^-$ state
to be in an SU(3) singlet state of $J^{PC}=1^{--}$. Therefore
the $0^-0^-$ final states through the gluon decay must be in an octet 
along the SU(3) breaking direction of $\lambda_8$. Since the leading 
term of the three-gluon decay is SU(3)-breaking, the one-photon 
process competes with the otherwise dominant gluon process, making 
it easier to determine a relative FSI phase through interference.

  The $J/\psi$ amplitudes are parametrized in terms of the  
reduced SU(3) amplitudes, $A_8$, $A_{\gamma}$, and $A_{\gamma 8}$,  
as follows:
\begin{equation}
     M(J/\psi\rightarrow 0^-0^-) =
 \sqrt{1/3}A_8{\rm tr}(P_8 P'_8 \lambda_8) + 
  A_{\gamma}{\rm tr}(P_8 P'_8 \lambda_{em}) \\
  + \sqrt{6}A_{\gamma 8}{\rm tr}(P_8\lambda_8 P'_8\lambda_{em}) 
    -(P_8\leftrightarrow P'_8),
   \label{parametrization}       
\end{equation}
where $P_8$ and $P_8'$ are the $3\times 3$ flavor
matrices of the $0^-$meson octet 
and $\lambda_{em}=(\lambda_3+\sqrt{1/3}\lambda_8$)/2. $A_8$ is for 
the gluon decay while $A_{\gamma}$ and $A_{\gamma 8}$ are for 
the one-photon annihilation and the SU(3) breaking correction to it, 
respectively.\footnote{The second order $\lambda_8$ breaking to the 
one-photon annihilation ${\rm tr}(P_8P'_8\lambda_8)
{\rm tr}(\lambda_8\lambda_{em})
-(P_8\leftrightarrow P'_8)$ has the same group structure as $A_8$.} 
No {\bf 10} or $\overline{\bf 10}$ representation of 
$0^-0^-$ arises from multiple insertions of $\lambda_8$ alone.
Charge conjugation invariance amounts to antisymmetrization in 
$P_8\leftrightarrow P'_8$, which forbids the {\bf 27}-representation
of $0^-0^-$. We have normalized each reduced amplitude such that 
sum of individual amplitudes squared be common. The decay amplitudes 
for the observed modes are listed in Table 1 in this parametrization. 
Also listed are the absolute values of the measured amplitudes\cite{PDG} 
after small phase-space corrections are made.
If the flavor SU(3) is a decent symmetry, $A_{\gamma 8}$ must be a 
fraction of $A_{\gamma}$. Knowing the magnitude of typical flavor-SU(3) 
breakings, let us allow 
\begin{equation}
        |A_{\gamma 8}| \leq 0.3\times|A_{\gamma}|. \label{expectation}
\end{equation} 

\section{Fits}

The one-photon annihilation amplitudes $A_{\gamma}$ and $A_{\gamma 8}$  
describe the electromagnetic form factors too. We have some theoretical 
understanding of their asymptotic behaviors. According to the perturbative 
QCD analysis\cite{farrar,brodsky}, the leading asymptotic behavior of the form 
factor for meson $M (=\pi^+, K^+)$ is given by
\begin{equation}
       F_M(q^2)\rightarrow\frac{16\pi\alpha_s(q^2)f_M^2}{-q^2}
   \biggl(1+\sum_{i=1}^{\infty}\frac{c_i}{[\ln(-q^2)]^{\gamma_i}} \biggr),  
          \label{asym}
\end{equation}   
where $f_M$ is the decay constant of $M$, $\alpha_s(q^2)$ is the QCD 
coupling, and $\gamma_i$ are positive constants. $F_M(q^2)$ approaches
a real value as $q^2 = m_{J/\psi}^2\rightarrow\infty$. Therefore, the
one-photon amplitudes $A_{\gamma}$ and $A_{\gamma 8}$ have a common phase
(=0) in this limit:
\begin{equation}
              \arg A_{\gamma 8}=\arg A_{\gamma}. \label{arg}
\end{equation}
Since $f_K\simeq 1.22\times f_{\pi}$, we expect $F_K(q^2) > F_{\pi}(q^2)$. 
The physical picture of this inequality is obvious in the spacelike 
region of $q^2$: Difference between $K^+$ and $\pi^+$ is $\overline{d}$ 
{\it vs} $\overline{s}$ in the valence quark content. 
The $\overline{s}$ quark, being a little heavier, is harder in momentum 
distribution inside $K^+$ than the $\overline{d}$ quark is inside
$\pi^+$. This leads to a stiffer form factor for $K^+$ than for $\pi^+$,
though an accurate theoretical estimate of $F_K(q^2)/F_{\pi}(q^2)$ is
not possible for finite $q^2$. With $f_K/f_{\pi}=1.22$, 
\begin{equation}
      \frac{A_{\gamma} + A_{\gamma 8}}{A_{\gamma} - A_{\gamma 8}} 
               \rightarrow 1.5 \label{ratio}
\end{equation}
as $m_{J/\psi}\rightarrow\infty$.
Combining Eq.(\ref{ratio}) with Eq.(\ref{arg}), we obtain $A_{\gamma 8}
\rightarrow 0.2\times A_{\gamma}$. When we keep the nonleading logarithmic 
terms, there is a small relative phase between $A_{\gamma}$ and $A_{\gamma 8}$;
\begin{equation}
              \arg(A_{\gamma 8}A_{\gamma}^*) = O\biggl(
 \frac{\gamma_1\pi}{[\ln(q^2/\Lambda_{QCD}^2)]^{1+\gamma_1}}\biggr).
\end{equation}
We may ignore it since it can be treated as a correction to 
the symmetry-breaking correction term $A_{\gamma 8}$.

If it happens that vector resonances of light quarks exist just 
around the $J/\psi$ mass, the form factors would not be asymptotic at this 
energy. If such high mass resonances should have a substantial branching into 
the $0^-0^-$ channels, the nonleading logarithmic terms would add up to a 
nonnegligible magnitude in Eq.(\ref{asym}).  In this case the phases of 
$A_{\gamma}$ and $A_{\gamma 8}$ would not be small. One may wonder
about whether a mass splitting of the resonances might generate a large 
phase difference between $A_{\gamma 8}$ and $A_{\gamma}$. However, the 
widths of such resonances, if any, would be so broad at such high mass 
that the mass splitting effect would be largely washed out.\footnote{ 
A glueball would have no effect 
on the phase difference between $A_{\gamma 8}$ and $A_{\gamma}$.}
Therefore we expect that the phase equality of Eq.(\ref{arg}) should 
hold in a good approximation. In our numerical analysis we shall set the 
phases of $A_{\gamma}$ and $A_{\gamma 8}$ to a common value and impose 
the condition of $F_K(q^2) \geq F_{\pi}(q^2)$ at $m_{J/\psi}$: 
\begin{equation}
           A_{\gamma 8}/A_{\gamma} \geq 0. \label{constraint}
\end{equation}

\subsection{Fit without FSI phases}
   If we attempt to fit the data with the leading terms $A_8$ and 
$A_{\gamma}$ alone without FSI phases, the result is unacceptable.
The fit of the minimum $\chi^2$ is obtained for $A_8 = 0.812$
and $A_{\gamma} = 0.807$ leading to $\chi^2 = 17.6$ for only three data.

   We then include $A_{\gamma 8}$ to fit the data. If we ignored the
constraint of Eq.(\ref{constraint}), the amplitudes could be fitted with
\begin{equation}
           A_8 = 0.739, \; A_{\gamma}= 0.814, \; A_{\gamma 8} = -0.228.
\end{equation}
This set of numbers would give $F_K(q^2)/F_{\pi}(q^2) = 0.63$ contrary 
to $F_K(q^2)/F_{\pi}(q^2)\geq 1$. When we include the constraint
$A_{\gamma 8}/A_{\gamma}\geq 0$, the fit of the best $\chi^2$ is back  
to $A_{\gamma 8}=0$ of $\chi^2 = 17.6$. The same poor fit with $A_8$
and $A_{\gamma}$ alone. It is fairly obvious why we cannot fit the data.
Looking up the parametrization in Table I, 
we see that without phases the $K^+K^-$ amplitude 
must be larger in magnitude than sum of the $\pi^+\pi^-$ and the 
$K^0\overline{K}^0$ amplitude for $A_{\gamma 8}/A_{\gamma}>0$. The
measured values badly violate this inequality. 

\subsection{Fit with $A_8$ and $A_{\gamma}$ including FSI phases}

  The natural recourse is to introduce FSI phases for the amplitudes. We 
first try with $A_8$ and $A_{\gamma}$ alone.  Defining the relative FSI
phase between $A_8$ and $A_{\gamma}$ by
\begin{equation}
         A_{\gamma} = e^{i\Delta} A_8,
\end{equation}
we can fit the amplitudes with
\begin{equation}
         \Delta = 89.6^{\circ}\pm 9.9^{\circ},\label{thefit}
\end{equation}
where $\Delta$ is defined between $0^{\circ}$ and $180^{\circ}$.
The attached uncertainty comes from the experimental errors of the 
branching fractions, which are treated as uncorrelated here. Since we
determine $\Delta$ through $\cos\Delta$ which is sensitive to small 
experimental errors near $\Delta=90^{\circ}$, the uncertainty in
Eq.(\ref{thefit}) turns out to be a little larger than one might expect from 
those of the branching 
fractions. One may wonder how much $\Delta$ can be reduced by 
adding the breaking term $A_{\gamma 8}$ with the constraint of 
Eq.(\ref{constraint}). The result is plotted in Fig.1. Dependence of 
$\Delta$ on the ratio $r = A_{\gamma 8}/A_{\gamma}$ is very mild: $\Delta$ 
decreases slowly and monotonically from $90^{\circ}$ at $r=0$ to 
$58^{\circ}$ at the edge of the allowed range, $r = 0.3$.  Even if 
the FSI phases of $A_{\gamma}$ and $A_{\gamma 8}$ are left independent,
it is fairly obvious that we cannot fit the data with small values for
all phase differences under the constraint 
$F_K(m^2_{J/\psi}) \geq F_{\pi}(m^2_{J/\psi})$. 
We have thus come to the conclusion that the FSI phase difference 
between the one-photon and the gluon decay amplitudes is very large, 
as large as $90^{\circ}$.  For this magnitude, 
it must come mostly from the long-distance FSI.  

\section{Perspectives}
Our conclusion of large FSI phases has a profound implication in the $B$ 
decay. The important input leading to this conclusion is that the 
electromagnetic form factor of $K^+$ does not fall faster than that of 
$\pi^+$. While it is very reasonable in perturbative QCD, we can 
in principle test this postulate in experiment. Just measure the ratio 
of the one-photon annihilation cross sections for
$e^+e^-\rightarrow \pi^+\pi^-$ and $K^+K^-$ off the 
$J/\psi$ peak. We do not have good data on the ratio 
$\sigma_{K^+K^-}/\sigma_{\pi^+\pi^-}$ off the peak. Experiment 
requires time and a good $\pi^+/K^+$ separation.
A measurement will certainly have a great impact on the issue of the 
long-distance FSI in heavy particle decays. The magnitude of measured
cross sections will also tell how close the form factors are to their 
asymptotic limits and therefore how small the phases of $A_{\gamma}$ 
and $A_{\gamma 8}$ are. Even a value of the unseparated ratio 
$\sigma_{K^0\overline{K}^0}/(\sigma_{K^+K^-}+\sigma_{\pi^+\pi^-})$ 
off the peak will throw in one more input in the analysis.

\acknowledgements
   I am grateful to S. Brodsky for an instruction in the 
perturbative OCD analysis of the form factors.  This work was 
supported in part by the Director, Office of Energy
Research, Office of High Energy and Nuclear Physics, Division of High
Energy Physics of the U.S. Department of Energy under Contract
DE--AC03--76SF00098 and in part by the National Science Foundation under
Grant PHY--95--14797.

\begin{table}
\caption{The SU(3) parametrization of the $0^-0^-$ decay amplitudes of $J/\psi$
and their magnitudes from the observed decay branching fractions. 
The central value of the $\pi^+\pi^-$ amplitude is normalized to unity.}
\begin{tabular}{cccc} 
Decay modes & $\pi^+\pi^-$ & $K^+K^-$ & $K^0\overline{K}^0$ \\ \hline
Parametrization & $\;\;\;A_{\gamma}-\sqrt{2/3}A_{8\gamma}$   
                & $A_8 +A_{\gamma} +\sqrt{2/3}A_{8\gamma}$ 
                & $A_8 -\sqrt{2/3}A_{8\gamma}$  \\ 
$|$Measured$|$      &1.000$\pm$0.078 & 1.367$\pm$0.089 & 0.925$\pm$0.060 
\end{tabular}
\label{table:1}
\end{table}


\noindent
\begin{figure}
\epsfig{file=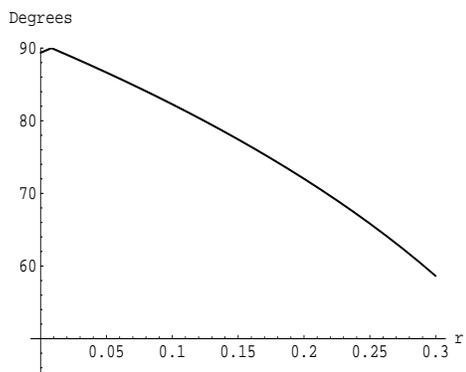,width=7cm,height=6cm}
\caption{The relative phase $\Delta$ in degrees between $A_8$ and 
$A_{\gamma}$ as a function of real parameter $r=A_{\gamma 8}/A_{\gamma}$ 
in its allowed range. $\Delta$ is drawn between $0^{\circ}$ and
$90^{\circ}$ since $\Delta\rightarrow 180^{\circ}-\Delta$ 
under the redefinition of $A_{\gamma}\rightarrow -A_{\gamma}$. 
\label{fig:1}} 
\end{figure}
 
\end{document}